\documentclass{PoS}
\usepackage{amsmath}
\usepackage{amssymb}
\usepackage{color}

\def\lsim {~^{<~}_{\sim~}}

\title{Two-dimensional phase structure of SU(2) gauge-Higgs model}

\ShortTitle{Two-dimensional phase structure of SU(2) gauge-Higgs model}

\author{\speaker{Shinya Gongyo}\\
        Department of Physics, Kyoto University\\
        E-mail: \email{gongyo@ruby.scphys.kyoto-u.ac.jp}}

\author{Daniel~Zwanziger\\
Department of Physics, New York University\\
E-mail: \email{dz2@nyu.edu}
}

\abstract{We study the phase structure of SU(2) gauge-Higgs model in two dimensions using lattice simulations. We show the result for the plaquette expectation value, static potential, and W propagator. 
Our results suggest that a confinement-like region and a Higgs-like region appear even in two dimensions. The behavior of the plaquette expectation value is consistent with a smooth cross-over in accordance with the Fradkin-Shenker-Osterwalder-Seiler theorem. In the confinement-like region, the static potential seems to rise linearly with string breaking at large distances, while in the Higgs-like region there seems to be a massive behavior which means that the BEH mechanism occurs. The correlation length obtained from the W propagator has a finite maximum between these phases, which supports no second-order phase transition.  Based on these results, we suggest that there is no phase transition in two dimensions.
}

\FullConference{The 32nd International Symposium on Lattice Field Theory\\
                 23-28 June, 2014\\
                 Columbia University New York, NY}

\begin{document}
\section{Introduction}

The Brout-Englert-Higgs mechanism (BEH mechanism), where the gauge boson becomes massive due to gauge symmetry breaking, is a common phenomenon extending from particle physics to condensed matter physics \cite{Higgs:1964ia}. In a superconductor, the Abelian BEH mechanism, known as the Meissner effect, occurs and in the standard model, the non-Abelian Higgs mechanism occurs.

The Higgs phenomenon on lattice has also been studied from the perspective of the confinement-Higgs behavior \cite{Fradkin:1978dv,Osterwalder:1977pc,Lang:1981qg,Caudy:2007sf,Montvay:1984wy,Bonati:2009pf,Maas:2010nc,Gongyo:2014jfa}. In the SU(2) gauge-Higgs model, the confinement-like region and the Higgs-like region appear in four dimensions, which may be connected analytically according to the Fradkin-Shenker-Osterwalder-Seiler theorem \cite{Fradkin:1978dv,Osterwalder:1977pc}. A characteristic behavior in the confinement-like region is that the static potential between the colored charges rises linearly until string breaking by pair production, while in the Higgs-like region it is of Yukawa type with a massive gauge boson \cite{Montvay:1984wy,Knechtli:2000df}. 
 
The case of two dimensions is of interest from the theoretical viewpoint. Because of the Coleman theorem and the Hohenberg-Mermin-Wagner theorem \cite{Coleman:1973ci,Hohenberg:1967zz}, Nambu-Goldstone bosons  do not appear and phase transitions do not occur. However, in the Higgs case, the Nambu-Goldstone bosons are absorbed by the gauge bosons and thus there are no massless bosons and these theorems do not apply.

In this contribution, we investigate the phase structure of the SU(2) gauge-Higgs model in two dimensions numerically. We show the gauge-invariant quantities such as the plaquette expectation value, static potential and W-propagator. 

\section{SU(2) gauge-Higgs model on lattice}
The lattice action of the SU(2) gauge-Higgs model with the fixed length of the Higgs field is given by
\begin{align}
S= \beta\sum_P \left(1-\frac{1}{2}\mathrm{Tr} U_{P}\right)
- \frac{\gamma}{2}\sum _{\mu , x}\mathrm {Tr}\left [\phi ^\dagger (x) U_\mu (x) \phi (x + \hat{\mu}) \right ] \label{FHaction}
\end{align}
with $U_\mu(x) \in SU(2) (\mu =1,2)$ the link-variable for the gauge field, $U_P \in SU(2)$ the plaquette-variable, and $\phi (x) \in SU(2)$ the Higgs field of the frozen length. This action, with frozen length of the Higgs fields, is formally derived from the SU(2) Higgs-Kibble model,
\begin{align}
S_l &= \beta\sum_P \left(1-\frac{1}{2}\mathrm{Tr} U_{P}\right) - \frac{1}{2}\sum _{\mu , x}\left(\varphi ^\dagger (x) U_\mu (x) \varphi (x + \hat{\mu})  + c.c. \right) \nonumber + \sum _x \lambda' \left(\varphi ^\dagger (x) \varphi (x) -\gamma \right)^2,
\end{align}
with $\varphi (x) = {}^t(\varphi ^1 (x), \varphi ^2 (x))$ in the SU(2) fundamental representation. By rescaling $\varphi (x)$ by $\varphi(x)=\gamma^{1/2}\tilde{\varphi}(x)$, and taking the limit $\lambda ' \rightarrow \infty$ with fixed $\gamma$, this action formally reduces to Eq.(\ref{FHaction}) with the SU(2) matrix, 
\begin{align}
\phi (x) = \left(
\begin{array}{cc}
	\tilde{\varphi} ^{*2} (x) & \tilde{\varphi} ^1 (x) \\
	-\tilde{\varphi} ^{*1} (x) & \tilde{\varphi} ^2 (x) \\
\end{array}
\right) .
\end{align}
In these models, the BEH mechanism occurs at the classical level, although whether it really occurs or not depends on $\beta$ and $\gamma$ \cite{Fradkin:1978dv,Caudy:2007sf,Montvay:1984wy,Maas:2010nc}. 

\section{Plaquette expectation value}
\label{Pl}
We show the plaquette expectation value to investigate if a phase transition appear or not. The plaquette expectation value is given by
\begin{align}
\left< P \right> \equiv \frac{1}{2}\left<\mathrm{Tr} U_{P} \right>
\end{align}
with $U_P$ the plaquette-variable. 

\begin{figure}[h]
\begin{center}
\includegraphics[scale=0.5]{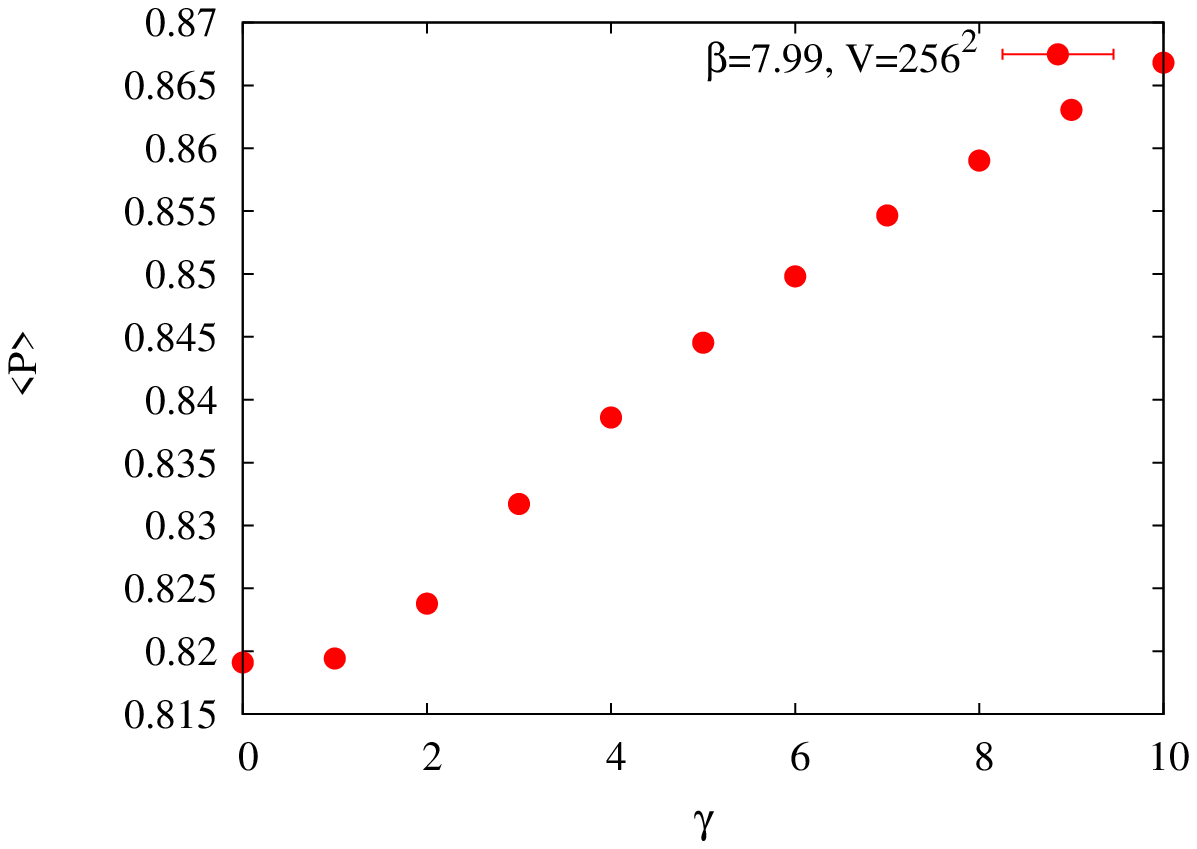}
\includegraphics[scale=0.5]{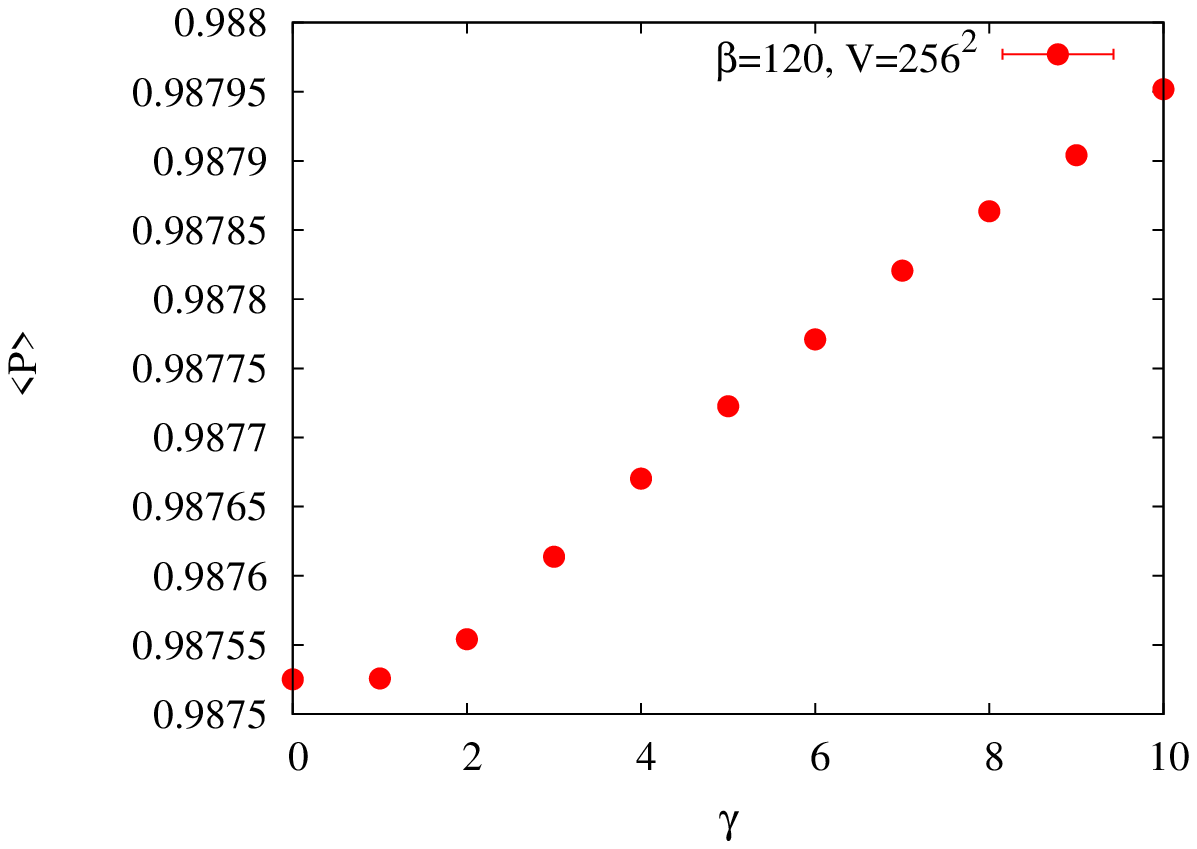}
\caption{\label{fig2.5}
The plaquette expectation value at $\beta = 7.99$ and $120$ on $256^2$}
\end{center}
\end{figure}

 In Fig.\ref{fig2.5}, we give the two-dimensional result at $\beta= 7.99$ and $120$ on $256^2$. The plaquette expectation values at both $\beta=7.99$ and $120$ increase smoothly with increasing $\gamma$ in the region $5 \lsim \gamma \lsim9$ in contrast to the gauge-dependent order parameter \cite{Gongyo:2014jfa}. We also investigated the volume-dependence for several values of $\gamma$ between $256^2$ and $512^2$ lattice and no difference was found. 

When we compare $\beta=7.99$ with $\beta =120$, the behavior of the plaquette expectation value seems to become smooth as $\beta$ increases and thus a transition might not appear in the region of $\beta >7.99$. In fact, if we take the limit of $\beta \rightarrow \infty$, we find $U_P \rightarrow 1$, corresponding to a pure gauge. In this case, after gauge transformation, the action corresponds to the two dimensional Heisenberg model, where a transition does not appear.

Furthermore, according to the Fradkin-Shenker-Osterwalder-Seiler theorem, no transition at small $\beta$ appears for gauge-invariant quantities. Therefore also in the region of $\beta < 7.99$, there might not be a transition, but a smooth cross-over. This is consistent with the recent work of Cubero and Orland \cite{Cubero:2014hla}, who find that there is no symmetry-breaking Higgs phase in the continuum theory.

\section{Static potential}
\label{Pot}
In two dimensional pure Yang-Mills theory, the Wilson-loop potential is exactly linear. On the other hand, if the BEH mechanism occurs, the potential behaves like
\begin{align}
V(r) &\sim -\int dp \frac{e^{-ipr}}{p^2+m^2}\sim -e^{-mr},
\end{align}
with $m$ the mass of vector boson, because the gauge boson becomes massive. 

\begin{figure}[h]
\begin{center}
\includegraphics[scale=0.5]{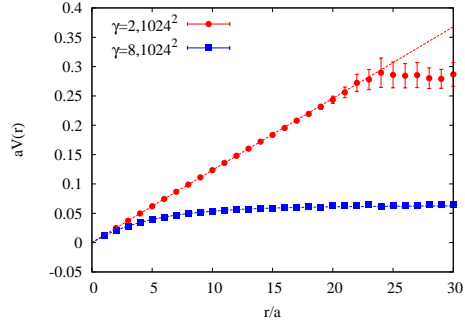}
\caption{
The static potential at $\beta=120$ and $\gamma = 2$ and $8$ in lattice units. The dashed lines show the fit results.}
\label{fig1}
\end{center}
\end{figure}

We calculated the static potential $V(r)$ using the Wilson loop on $1024^2$ lattice at $\beta = 120$ and $\gamma =2$ and $8$ as shown in FIG. \ref{fig1}. The static potential is fitted by the form,
\begin{align}
-Ae^{-mr}+ \sigma r + B, \label{fitform}
\end{align}
with the parameter $A,B,m$ and $\sigma$ in the region of $r/a <20$, and the results are summarized in Table \ref{TableI}.

In FIG. \ref{fig1}, the potential at $\gamma =2$ shows the linear potential and the string breaking at $r/a \sim 20$, while the potential of $\gamma=8$ does not show the linear potential, but behaves like $-e^{-mr}$. Thus $\gamma =2$ corresponds to confinement-like region and $\gamma =8$ corresponds to Higgs-like region. Furthermore because of string breaking, the difference between the regions may be quantitative (though strong), which indicates that these regions are connected analytically.
\begin{table}[h]
\caption{ The fit results of the static potential by $-Ae^{-mr}+ \sigma r + B$ for $r/a<20$. }
\label{TableI}
\begin{center} 
\begin{tabular}{cccccc}
\hline
\hline
  $\gamma$     &$ m$ &  $\sigma$ & $A$ & $B$ & $\chi/N$ \\
\hline
$ 2$                &    $0.19(7)$    &  $0.01221(5)$ & $0.0015(7)$ & $0.0015(7)$&$0.3$\\
$8$                 &     $0.207(4)$   &  $7(9)  \times 10^{-5}$ &  $ 0.061(1) $ & $ 0.060 (1) $ &$1.2$\\
\hline
\hline
\end{tabular}
\end{center} 
\end{table}

The fit results support the appearance of these regions. At $\gamma =2$ a finite string tension appears, which indicates the confinement-like region. Though the mass parameter $m$ does not seem to be small, the prefactor $A$ is small. Therefore the first term in Eq. (\ref{fitform}) is negligible compared with the second term $\sigma r$. At $\gamma=8$, the string tension $\sigma$ is almost $0$, which means that the potential does not show a linear slope. 

\section{W propagator}
\label{Wp}
Finally, we show another gauge-invariant quantity, the W propagator, defined by 
\begin{align}
D_{\mu \nu}(x-y) = \frac{1}{3}\sum_{a=1,2,3}\left< W_\mu ^a (x) W_\nu ^a (y)\right>,
\end{align}
where 
\begin{align}
 W_\mu ^a (x) \equiv \frac{1}{i}\mathrm{Tr} \left[\tau ^a \left\{\phi ^\dagger (x)U_\mu (x)\phi (x + \hat{\mu} )\right\}\right] \label{Wfield}
\end{align}
with $\tau ^a$ SU(2) generator. Note that the W field is gauge invariant and the effective mass for the transverse part of the W field corresponds to the two-dimensional analog of a $1^- $ state \cite{Montvay:1984wy, Maas:2010nc}, which is a singlet state with negative parity. This propagator coincides with the gluon propagator  in the unitary gauge, $\phi (x) =1$ and thus it is easily calculated.

The effective mass of the W field is estimated from the linear slope of the logarithm of the propagator at zero-spatial-momentum,
\begin{align}
D^0_{\mu \nu}(t) =  \frac{1}{3V}\sum_{a,x_1,y_1}\left< W_\mu ^a (x_1,t) W_\nu ^a (y_1,0)\right>
\end{align}
with the two dimensional volume $V$.
The  effective mass of the transverse part is given by $D^0_{11}(t)$ and that of the longitudinal part is given by $D^0_{22}(t)$ \cite{Gongyo:2013sha}. In this contribution, we estimated only the transverse effective mass.
\begin{figure}[h]
\begin{center}
\includegraphics[scale=0.7]{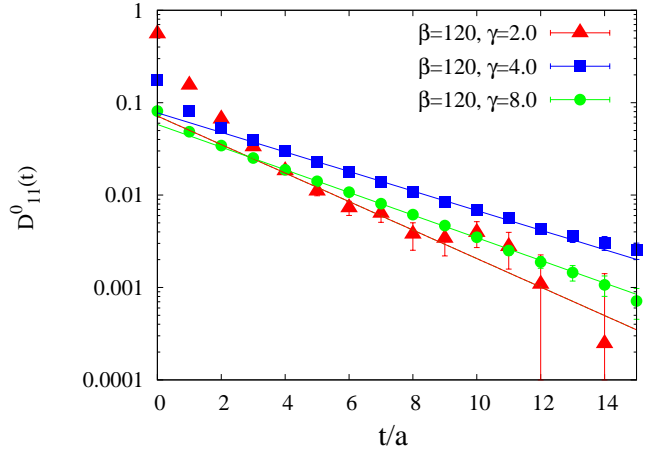}
\caption{\label{fig2}
The propagator at zero-spatial-momentum, $D^0_{11}(t)$, on $256^2$ at $\beta = 120$ and $\gamma = 2,4$ and $8$ in lattice units.}
\end{center}
\end{figure}

We show $D^0_{11}(t)$ on $256^2$ lattice at $\beta = 120$ and $\gamma =2,4$ and $8$ in FIG. \ref{fig2}. The logarithm of these zero-spatial-momentum propagators seems to be almost linear or convex upwards, which indicates the preservation of the Kallen-Lehmann representation, in contrast with the gluon propagator in other gauges like the Landau gauge and MA gauge \cite{Mandula:1987rh}. In particular, the two dimensional gluon propagator in the Landau gauge shows remarkable violation of the Kallen-Lehmann representation and goes to zero at zero momentum \cite{Gongyo:2014jfa,Maas:2007uv,Cucchieri:2007rg,Zwanziger:2012xg}.
\begin{table}[h]
\caption{ The fit results of the zero-spatial-momentum propagator by $A_D e^{-m_D t}$ for $t/a=4-15$ at $\beta =120$ and on $256^2$ and $512^2$. 
}
\label{TableII}
\begin{center} 
\begin{tabular}{ccccc}
\hline
\hline
  $\gamma$     &$ m_D$ & $A_D$ & $L^2$& $\chi/N$ \\
\hline
$ 2$                &  $0.36(3)$ & $0.07(1)$&$256$&$0.8$\\
$4$                 &  $ 0.243(4) $ & $ 0.078 (1) $ &$256$&$0.5$\\
$5$                 &  $ 0.240(2) $ & $ 0.069 (1) $ &$256$&$0.1$\\
$ 6$                &  $0.265(5)$ & $0.069(2)$&$256$&$0.9$\\
$8$                 &  $ 0.282(1) $ & $ 0.058 (1) $ &$256$&$0.08$\\
$10$                 &  $ 0.304(2) $ & $ 0.051 (1) $ &$256$&$0.2$\\
$4$                 &  $ 0.243(2) $ & $ 0.077 (1) $ &$512$&$1.0$\\
$5$                 &  $ 0.237(3) $ & $ 0.067 (1) $ &$512$&$0.6$\\
$ 6$                &  $0.267(2)$ & $0.067(1)$&$512$&$1.0$\\
\hline
\hline
\end{tabular}
\end{center} 
\end{table}

In TABLE. \ref{TableII}, we summarize for $D^0_{11}(t)$ the fit result of $A_D e^{-m_D t}$ with the parameters $A_D$ and $m_D$ for $t/a=4-15$. The mass parameter $m_D$ in lattice units decreases in the region of $\gamma<5$, and increases in the region of $\gamma>5$. In other words the mass parameter shows minimum at $\gamma \simeq 5$, which means that the correlation is maximum. We also show the volume-dependence of the mass parameters in TABLE. \ref{TableII} and there seems to be almost no volume-dependence between $256^2$ and $512^2$ at $\gamma=4,5$ and $6$. The minimum does not seem to go to zero with increasing volume. Therefore it is not a second-order phase transition, consistent with the result of the praquette expectation value.

\section{Summary}
We have shown the numerical result of the two dimensional phase structure in SU(2) gauge-Higgs model.
We have carried out the Monte Carlo simulation and calculated gauge-invariant quantities such as the plaquette expectation value at $\beta=7.99$ and $120$, static potential and the W propagator at $\beta = 120$.

The plaquette expectation value shows a smooth cross-over between confinement-like region and a Higgs-like region at $\beta=7.99$ and $120$ which is consistent with the Fradkin-Shenker-Ostervalder-Seiler theorem. It indicates no phase transition in two dimensions.
The static potential shows a linear rise and string breaking at $\gamma=2,$ and behaves like a two-dimensional Yukawa potential at $\gamma =8$. The effective mass of the W propagator in lattice units has a minimum at $\gamma \simeq 5$ which does not go to zero with increasing volume. These results support that there is a BEH mechanism in the region of $\gamma > 5$ and a confinement-like region and a Higgs-like region appear even in two dimensions. 

\section*{Acknowledgements}
The authors are grateful to A.~Maas for suggesting this investigation. They thank J.~Greensite, P.~Hohenberg, A.~Maas, M.~Schaden and A.~Sokal for discussions and comments. 
S.~G.~is supported by a Grant-in-Aid for Scientific Research from the JSPS Fellows (No. 24-1458).   
The lattice calculations were done on NEC SX-9 at Osaka University.


\begin{thebibliography}{99}
\bibitem{Higgs:1964ia} 
  P.~W.~Anderson,
  Phys.\ Rev.\  {\bf 130}, 439 (1963);
  P.~W.~Higgs,
  Phys.\ Lett.\  {\bf 12}, 132 (1964);
  Phys.\ Rev.\ Lett.\  {\bf 13}, 508 (1964);
  F.~Englert and R.~Brout,
  Phys.\ Rev.\ Lett.\  {\bf 13}, 321 (1964).

\bibitem{Fradkin:1978dv} 
  E.~H.~Fradkin and S.~H.~Shenker,
  Phys.\ Rev.\ D {\bf 19}, 3682 (1979).
 
\bibitem{Osterwalder:1977pc} 
  K.~Osterwalder and E.~Seiler,
  Annals Phys.\  {\bf 110}, 440 (1978).

\bibitem{Lang:1981qg} 
  C.~B.~Lang, C.~Rebbi and M.~Virasoro,
  Phys.\ Lett.\ B {\bf 104}, 294 (1981);
 
  J.~Jersak, C.~B.~Lang, T.~Neuhaus and G.~Vones,
  Phys.\ Rev.\ D {\bf 32}, 2761 (1985).

\bibitem{Caudy:2007sf} 
  W.~Caudy and J.~Greensite,
  Phys.\ Rev.\ D {\bf 78}, 025018 (2008)
  [arXiv:0712.0999 [hep-lat]];
  J.~Greensite, S.~Olejnik and D.~Zwanziger,
  Phys.\ Rev.\ D {\bf 69}, 074506 (2004)
  [hep-lat/0401003].

\bibitem{Montvay:1984wy} 
  I.~Montvay,
  Phys.\ Lett.\ B {\bf 150}, 441 (1985);
  Nucl.\ Phys.\ B {\bf 269}, 170 (1986);
  W.~Langguth, I.~Montvay and P.~Weisz,
  Nucl.\ Phys.\ B {\bf 277}, 11 (1986).

\bibitem{Bonati:2009pf} 
  C.~Bonati, G.~Cossu, M.~D'Elia and A.~Di Giacomo,
  Nucl.\ Phys.\ B {\bf 828}, 390 (2010)
  [arXiv:0911.1721 [hep-lat]].

\bibitem{Maas:2010nc} 
  A.~Maas,
  Eur.\ Phys.\ J.\ C {\bf 71}, 1548 (2011)
  [arXiv:1007.0729 [hep-lat]];
  Mod.\ Phys.\ Lett.\ A {\bf 28}, 1350103 (2013)
  [arXiv:1205.6625 [hep-lat]];
  A.~Maas and T.~Mufti,
  arXiv:1312.4873 [hep-lat].

\bibitem{Gongyo:2014jfa} 
  S.~Gongyo and D.~Zwanziger,
  arXiv:1402.7124 [hep-lat].

\bibitem{Knechtli:2000df} 
  F.~Knechtli {\it et al.}  [ALPHA Collaboration],
  Nucl.\ Phys.\ B {\bf 590}, 309 (2000)
  [hep-lat/0005021].

\bibitem{Coleman:1973ci} 
  S.~R.~Coleman,
  Commun.\ Math.\ Phys.\  {\bf 31}, 259 (1973).

\bibitem{Hohenberg:1967zz} 
  P.~C.~Hohenberg,
  Phys.\ Rev.\  {\bf 158}, 383 (1967);
 
  N.~D.~Mermin and H.~Wagner,
  Phys.\ Rev.\ Lett.\  {\bf 17}, 1133 (1966).

\bibitem{Maas:2011se} 
  A.~Maas,
  Phys.\ Rept.\  {\bf 524}, 203 (2013)
  [arXiv:1106.3942 [hep-ph]], and references therein.

\bibitem{Bogolubsky:2009dc} 
  I.~L.~Bogolubsky, E.~M.~Ilgenfritz, M.~Muller-Preussker and A.~Sternbeck,
  Phys.\ Lett.\ B {\bf 676}, 69 (2009)
  [arXiv:0901.0736 [hep-lat]].

\bibitem{Dudal:2010tf} 
  D.~Dudal, O.~Oliveira and N.~Vandersickel,
  Phys.\ Rev.\ D {\bf 81}, 074505 (2010)
  [arXiv:1002.2374 [hep-lat]].

\bibitem{Oliveira:2012eh} 
  O.~Oliveira and P.~J.~Silva,
  Phys.\ Rev.\ D {\bf 86}, 114513 (2012)
  [arXiv:1207.3029 [hep-lat]].

\bibitem{Maas:2007uv} 
  A.~Maas,
  Phys.\ Rev.\ D {\bf 75}, 116004 (2007)
  [arXiv:0704.0722 [hep-lat]].

\bibitem{Cucchieri:2007rg} 
  A.~Cucchieri and T.~Mendes,
  Phys.\ Rev.\ Lett.\  {\bf 100}, 241601 (2008)
  [arXiv:0712.3517 [hep-lat]];
  Phys.\ Rev.\ D {\bf 78}, 094503 (2008)
  [arXiv:0804.2371 [hep-lat]].

\bibitem{Gongyo:2013sha} 
  S.~Gongyo and H.~Suganuma,
  Phys.\ Rev.\ D {\bf 87}, 074506 (2013)
  [arXiv:1302.6181 [hep-lat]];
  S.~Gongyo, T.~Iritani and H.~Suganuma,
  Phys.\ Rev.\ D {\bf 86}, 094018 (2012)
  [arXiv:1207.4377 [hep-lat]].
  
\bibitem{Mandula:1987rh} 
  J.~E.~Mandula and M.~Ogilvie,
  Phys.\ Lett.\ B {\bf 185}, 127 (1987).

\bibitem{Maas:2013sca} 
  A.~Maas and D.~Zwanziger,
  Phys.\ Rev.\ D {\bf 89}, 034011 (2014)
  [arXiv:1309.1957 [hep-lat]].

\bibitem{Cubero:2014hla} 
  A.~C.~Cubero and P.~Orland,
  Phys.\ Rev.\ D {\bf 89}, 085027 (2014)
  [arXiv:1403.0276 [hep-th]].

\bibitem{Zwanziger:2012xg} 
  D.~Zwanziger,
  Phys.\ Rev.\ D {\bf 87}, 085039 (2013)
  [arXiv:1209.1974 [hep-ph]].
 
\end{thebibliography}
\end{document}